\documentclass[preprint]{aastex}
\usepackage{graphicx}
\def\edcomment#1{\iffalse\marginpar{\raggedright\sl#1\/}\else\relax\fi}
\reversemarginpar

\begin{document}
\title{Probing Interstellar Dust Models Through SAXS \\ (Small Angle X-Ray Scattering)}
 \author{Eli Dwek and Viktor Zubko}
\affil{NASA Goddard Space Flight Center, Code 685, Greenbelt, MD 20771}
\author{Richard G. Arendt}
\affil{SSAI, NASA Goddard Space Flight Center, Code 685, Greenbelt, MD 20771}
\author{Randall K. Smith}
\affil{Smithsonian Astrophysical Observatory, 60 Garden St., Cambridge, MA 02138}

\begin{abstract}
 A viable interstellar dust model - characterized by  the composition, morphology, and size distribution of the dust grains and by the abundance of the different elements locked up in the dust -  should fit all observational constraints  arising primarily from the interactions of the dust with incident radiation or the ambient gas. 
In spite of the many different manifestations of these interactions,
we still lack a comprehensive dust model that is consistent with {\it all} the observational
constraints. 

An important advance towards the construction of such a model was recently made by Zubko, Dwek, and Arendt (2003, ZDA03) who, for the first time, included the average interstellar extinction, the diffuse infrared emission, {\it and} the interstellar abundances as explicit constraints in the construction of models consisting of astrophysical dust particles without any predetermined functional form for their grain size distribution. The results showed the existence of many distinct dust models that satisfy a basic set of observational constraints. 

X-ray halos, produced primarily by small angle scattering off large dust particles along the line of sight to bright X-ray sources, probe dust properties largely inaccessible at other wavelengths. In this contribution we briefly review the microscopic physics and the macroscopic effects that determine the intensity and spatial profile of X-ray halos. Focusing on the X-ray halo around the bright source GX~13+1, we show that halo observations can be used to discriminate between the different, currently viable, ZDA03 dust models.  X-ray halos may therefore be an essential constraint that need to be explicitly included in all future dust models.
\end{abstract}

\section{Introduction}

The presence of dust in the general interstellar medium (ISM) is primarily inferred from the absorption, scattering and polarization of
stellar and diffuse Galactic light, the infrared, submillimeter, and microwave emission spectrum from the diffuse ISM,
the presence of broad solid state emission and absorption features in stellar and nebular spectra, and from the presence of X-ray scattering halos around X-ray sources.
 Indirect evidence for the presence of dust is inferred from the depletion of
various refractory elements from the gas phase of the ISM, the presence of "presolar" dust grains
in the solar system, and the role of dust as a catalyst for chemical reactions, and as a provider of photoelectrons in the heating of the neutral ISM. 
Comprising less than one percent of the mass of the interstellar gas, dust nevertheless plays a major role in the radiative appearance of galaxies, and the thermal and chemical balance of their interstellar medium.

Ideally, an interstellar dust model should be derived by {\it simultaneously} fitting all observational constraints such as  the average interstellar extinction, the diffuse IR emission, the UV to optical wavelength dependent albedo, polarization, and scattering phase function, and the intensity and spatial distribution of X-ray halos. It must consist of particles with known optical, physical, and chemical properties, and require no more than the ISM abundance of any given element to be locked up in the dust. In practice, most interstellar dust models have been constructed by deriving the abundances and size distributions of some well studied  solids, such as graphite or silicates, from select observations such as the average interstellar extinction, the polarization, or the diffuse infrared emission, and checked for consistency with other observational constraints such as the wavelength dependent albedo and interstellar abundances.

The first widely used dust model derived in such manner consisted of spherical, bare silicate and graphite particles characterized by a ${\rm d}n/{\rm d}a \propto a^{-3.5}$ power law distribution in grain radius $a$ in, respectively, the 0.025 to 0.25 $\mu$m, and  the 0.005 to 0.25 $\micron$ size intervals (Mathis, Rumple, \& Nordsieck 1977; MRN). 
Using the dust optical constants of Draine \& Lee (1984; DL84), the MRN model
provides a good fit to the average interstellar extinction curve. Assuming solar composition for the ISM, the model requires essentially all
the interstellar carbon, C/H = 370 ppm (parts per million), and all the magnesium,
silicon, and iron, Mg/H, Si/H, Fe/H = \{34, 35, 28\}, to be locked up in dust.

The shortcomings of deriving an interstellar dust model from only one set of observational constraints were demonstrated by the results of the {\it Infrared Astronomical Satellite} ({\it IRAS}) all sky survey,
which provided the average infrared (IR) emission spectrum  at 12, 25, 60, and 100 $\mu$m
from the diffuse ISM. The observations showed an excess of 12 and 25 $\mu$m emission over
that expected from dust heated by the local interstellar radiation field (ISRF) and
radiating at the equilibrium dust temperature. Draine \& Anderson (1985) suggested that
the MRN grain size distribution should be extended to very small grains (VSG) with
radii of $\sim$ 5 \AA\ which undergo temperature fluctuations when heated by the ISRF.
Allamandola, Tielens, \& Barker (1985), and L\'eger \& Puget (1984) identified these
VSG with polycyclic aromatic hydrocarbon (PAH) molecules whose presence in the ISM was
inferred from the ubiquitous solid state emission features at 3.3, 6.7, 7.6, 8.6, and
11.3 $\mu$m. Other possible candidates for the carriers of these IR emission features were summarized in a review by Tokunaga (1997). 

The first models that attempted to fit in a self--consistent
manner the interstellar extinction as well as the diffuse IR emission using PAHs as
an interstellar dust component were presented by D\'esert, Boulanger, \& Puget (1990) and Siebenmorgen \& Kr\"ugel (1992).
However, their models used an ad hoc formula to characterize the optical and UV extinction of the PAHs. 

Interstellar polarization provides additional constraints on interstellar dust particles
(Kim \& Martin 1995, 1996; Li \& Greenberg 1997). In particular, the latter authors
presented an interstellar dust model consisting of PAHs, and cylindrical silicates
coated with an organic refractory mantle. Their model satisfies the interstellar
extinction, polarization, and solar abundance constraints. It did not attempt to fit
the diffuse IR emission, and used particles with hypothetical optical properties
to represent the far-UV extinction and the 2175 {\AA} extinction bump. 

The Diffuse IR Background Experiment (DIRBE) and Far Infrared Absolute
Spectrophotometer (FIRAS) instrument on board the {\it Cosmic Background Explorer}
({\it COBE}) satellite provided the most extensive wavelength coverage
(3.5 to 1000 $\mu$m) of the IR emission from the diffuse ISM, and indirect evidence
for the emission from PAHs from this phase of the medium (Dwek et al. 1997).
Dwek et al. (1997) attempted to fit the interstellar extinction and diffuse IR emission
using a mix of bare silicate and graphite particles with DL84
optical constants, and PAHs with the UV and optical properties of D\'esert, Boulanger, \& Puget (1990).
The model failed to reproduce the observed interstellar extinction, primarily due
to the non-physical nature of the UV-optical properties adopted for the PAHs, and the adherence to a power law representation for the grain size distribution.
Draine \& Li (2001) and Li \& Draine (2001; LD01) improved on this type of model
by using a more realistic characterization of the optical properties of the PAHs,
based on laboratory measurements. However, they used a predetermined functional form for the grain size distribution (Weingartner \& Draine 2001; WD01), and did not include interstellar abundance constraints in the fitting procedure. Consequently, their model requires an excessive amount
of Mg, Si, and Fe to be locked up in dust. The Si/H = Mg/H = Fe/H abundance
in the dust is about 50 ppm (parts-per-million = 10$^{-6}$), almost twice
the amount of iron if the ISM has solar abundances.

Since interstellar abundances play a crucial role in constraining the amount
of matter that can be locked up in the dust, Zubko, Dwek, \& Arendt (2003; ZDA03) included
interstellar abundances as an explicit constraint
in the construction of dust models. Previously, models were just examined for consistency with abundance
constraints, and non-compliances were given little weight because of the uncertainties
in the determination of interstellar abundances. ZDA03 explored
a variety of ISM abundances and potential dust compositions. The models are physical in the sense that they use measured optical constants or  observed
radiative properties to characterize the optical properties of the various
dust constituents.

We first present, in \S2, a brief summary of the ZDA03 dust models. An important result of their investigation is that there exist many, distinct, models that simultaneously satisfy the extinction, IR emission, and abundance constraints. X-ray halos provide an important additional constraint on vaible interstellar dust models. They are produced by the scattering of X-rays primarily off large dust grains, a region of grain sizes largely inaccessible at other wavelengths. In \S3 we briefly summarize the basic X-ray scattering properties of individual dust grains, and the collective effects  of many grains along the line of sight to an X-ray source on the intensity and spatial profile of its scattering halo. We review recent modeling efforts of the halo around Nova Cygni 1992, and then present preliminary models for the halo around the  X-ray source GX13+1, showing that is a powerful discriminator between different classes of dust models, significantly narrowing the number of viable
models presented by ZDA03 (\S4). A brief summary of the paper is presented in \S5.

\section{Interstellar Dust Models Consistent With Extinction, IR Emission, and Abundance Constraints}

Given the composition and physical and optical properties of the dust particles, the problem
of simultaneously fitting the model to a set of observational constraints is
a typical ill-posed inversion problem, in which the grain size distribution
is the unknown. 
ZDA03 solved the inversion problem using the method of regularization, and considered five different dust compositions as potential model ingredients: (1) PAHs; (2) graphite; (3) hydrogenated amorphous
carbon of type ACH2; (4) silicates (MgSiFeO$_4$); and (5) composite particles
containing different proportions of silicates, organic refractory material (C$_8$H$_8$O$_4$N), water ice (H$_2$O), and voids.

\noindent
These five dust compositions were combined to create five different
classes of dust models. \\
{\bf The first class} consists of PAHs, and bare graphite and
silicate grains, and is identical to the carbonaceous/silicate model recently
proposed by LD01. \\
 {\bf The second class} of models contains composite particles in addition
to PAHs, bare graphite and silicate grains. \\
 {\bf The third and fourth classes} of models comprise the first and second
classes but with the graphite grains completely replaced by amorphous
carbon grains. \\
 {\bf In the fifth class} of models the only carbon is in PAHs and in the organic refractory material in composite grains. That is, the model comprises only PAHs, bare silicate, and composite particles. 

Models including only PAHs and bare grains were designated as BARE, and
those containing additional composite particles as COMP. 
The BARE and COMP models were further subdivided according to the composition
of the bare carbon particles used in the model: graphite, amorphous carbon,
and no carbon, designated respectively by -GR, -AC, and -NC. 
For all model classes, ZDA03 performed the fit using three different
sets of ISM abundances: solar, B~stars,
and F and G~ stars (correspondingly designated as -S, -B, and -FG submodels).
So, for example, a BARE model with graphite grains derived by assuming
the solar ISM abundances is designated as BARE-GR-S, or a COMP model
with no carbon obtained by assuming B star ISM abundances is
designated as COMP-NC-B.

\begin{figure}
\caption{Flow chart depicting the methodology and results of ZDA03, presented as a poster at this meeting.}
\end{figure}

The method and the results of the analysis are shown in the form of a work flow chart in the top half of Figure 1 which is an edited reproduction of the poster exhibited at this conference.

{\bf The left part of the {\it Work Flow Chart}} depicts the observational
constraints used in the work of ZDA03. They include the average
interstellar extinction, the diffuse IR emission, and the abundance
constraints.

{\bf The right part of the {\it Work Flow Chart}} summarizes the different
dust compositions used in the model.  

{\bf The center of the {\it Work Flow Chart}} depicts the method of
regularization, and the different classes of dust models used in the fit
to the observational constraints.

Model results are depicted below
the {\it Work Flow Chart}. Of particular interest are the curves in the upper
row, showing the derived grain size distributions for the different models.
The lowest row shows the abundances of the different elements locked up
in the dust. B stars abundances place the most stringent constraint
on the amount of carbon that can be locked up in the dust, less than
$\approx$ 160 ppm. A model that could fit this constraint
(as well as the extinction and IR emission) consisted of PAH, bare silicate,
and composite grains (no bare carbon particles!). In this model
(model COMP-NC-B) the 2175 \AA\ extinction bump was entirely
produced by PAHs. 

Figure 2 offers a detailed comparison of
the PAH and bare graphite and silicate model derived by
LD01 and ZDA03. The main peak in the LD01 graphite size
distribution peaks at larger grain sizes ($\sim 0.3 \mu$m),
whereas the ZDA03 peaks at $\sim 0.1 \mu$m. The ZDA03 size distribution
is smoother, only double--peaked: one peak corresponding to graphite grains, the other to PAHs. It lacks the more complex triple--peak distribution of the LD01 model.  Both models consume about equal amounts of carbon in PAHs
and graphite with C/H $\approx$ 240 ppm. The models differ significantly in the size distribution of the silicate particles. The LD01
silicate size distribution is steeper and has a larger amplitude,
whereas the ZDA03 distribution has a slope similar to the MRN size
distribution, a significantly lower peak, and a larger mass fraction
of small silicate grains. The amount of Si/H, Mg/H, and Fe/H in the ZDA03 model is about
33 ppm for each element, consistent with solar abundances, whereas
the LD01 model requires an abundance of $\sim$ 50 ppm
for each element, almost twice the available iron abundance if
the ISM had solar composition. 

\begin{figure}[h]
\begin{center}
\end{center}
\caption{Grain-size distributions for the BARE-GR models derived by Zubko, Dwek, \& Arendt (2003) are compared to that derived by Li \& Draine (2001) for the same dust composition.}
\end{figure}

 The results summarized above show that there are many classes of interstellar dust models that provide good simultaneous fits to the far-UV to near-IR extinction, thermal
   IR emission, and elemental abundance constraints. The models can be grouped into two major categories: BARE and COMP models. The latter are distinguished from the former by the fact that they contain a population of composite particles which generally comprise of larger grains than bare particles. Furthermore, a significant fraction of the volume of composite grains consists of voids, so their effective electron density is significantly smaller than that of the bare particles. Since X-ray halos are primarily produced by scattering off  electrons in large grains, they may provide an excellent discriminator between viable dust models.

\section{The Production of X-Ray Halos}

The use of X-ray halos as diagnostic tool for determining dust properties, the morphology of the ISM, and distances to X-ray sources, have been previously discussed in the literature by, among others, Overbeck (1965), Stecher \& Williams (1966), Tr\"umper \& Sch\"onfelder (1973), Alcock \& Hatchett (1978), Mauche \& Gorenstein (1986), Mathis \& Lee (1991), Predehl \& Schmitt (1995), Predehl \& Klose (1996), and Predehl et al. (2000). Here we will concentrate on the two main factors that determine the X-ray halo profile and intensity: (1) the scattering properties of the individual dust grains, and (2) the collective effects of the many individual dust grains along the line of sight (LOS) to the X-ray source.

\subsection{The Scattering of X-Rays by Single Dust Particles}

The scattering of X-rays is usually described by the Rayleigh-Gans theory which is valid under the following two assumptions  (van der Hulst 1981): \\
    (1) $|m-1| \ll 1$, where $m$ is the refractive index of the dust particle. This conditions ensures that the incident waves are not significantly reflected off the dust; and
   (2) $\Delta \Phi = (4\pi a/\lambda) |m-1| \ll 1$, ensuring that there is no appreciable phase shift across the grain, so that each element in the grain is exposed to the same phase and amplitude of the incident wave. 
   
 The second condition is the more stringent one which can be written as (Alcock \& Hatchett 1978):
 \begin{equation}
 E_x(keV)  \gg  3\ a(\micron)\ \left( {\rho \over 3\ {\rm g\ cm}^{-3}}\right) \qquad,
\end{equation}
where $E_x$ is energy of the incident X-rays. The Rayleigh--Gans approximation is therefore valid for X-ray energies in excess of a few keV, and dust particles with radii $a < 1\ \micron$.

When these conditions are met, each electron in the grain can be considered as an independent scatterer. Waves scattered in a given direction interfere with each other because of the different positions of the scatterers within the grain.  The scattering is strongly peaked in the forward direction since, when $\theta \ll \lambda/a$, the phase differences between the scattering electrons across the grain are very small. 

The interference effects can be readily calculated for simple grain geometries.
Consider an X-ray incident on a dust particle, which for illustrative purposes is taken to be a sphere of radius $a$ (see Figure 3). Since X-ray energies are in general much larger than the electron binding energy to an atom, the electrons can be considered as free particles, confined to the volume of the dust grain, and vibrating at the frequency of the incident X-ray. Let $N$ be the position in the dust of an electron scattering the X-ray at an angle $\theta$ with respect to the direction of the incident beam. The path difference between the scattered rays eminating from positions $N$ and $P$ is given by $\overline {OP} - \overline {MN} = \underline r \cdot (\underline k - \underline k_0) \equiv \underline {K} \cdot \underline {r}$, where $\underline k_0$ and $\underline k$ are unit vectors in the direction of the incident and scattered beam, respectively. From Fig. 3 we see that $|\underline K| = 2 \sin(\theta/2)$. \\ The phase difference between the two rays is therefore given by 
\begin{eqnarray}
\Phi & = & {2 \pi \over \lambda}\ \underline K \cdot \underline r \\ \nonumber 
 & = & {4 \pi r \over \lambda}  \sin\left({\theta\over 2}\right)\ \cos\alpha
\end{eqnarray}
where $\alpha$ is the angle between $\underline r$ and $\underline K$. 
The scattering is coherent, so the amplitudes are first added and then squared to get the intensity of the scattered light. The scattering factor, defined as the normalized sum of the scattering amplitudes in a given direction, is given by:
\begin{eqnarray}
f(\theta)  & =  & \int \rho(r)\ \exp(i\Phi)\ {\rm d}V \\ \nonumber
& =  & \int \rho(r)\ \exp\left[i \kappa r \cos\alpha \right]\ 2 \pi r^2 {\rm d}r \sin\alpha{\rm d} \alpha
\end{eqnarray}
where $\kappa \equiv 4 \pi \sin(\theta/2)/\lambda$, d$V  =  2\pi r^2 $d$r\sin\alpha$d$ \alpha$ is the differential volume element in a spherical coordinate system in which the $z-$axis is in the $\underline K$ direction, and $\rho(r)$ is the electron density in the solid.
For a homogeneous dust particle the scattering factor simplifies to (see van de Hulst 1981, \S7.21):
\begin{eqnarray}
f(\theta) & =  & \int 4 \pi \rho(r) r^2 \ {\sin (\kappa r)\over \kappa r} {\rm d}r \\ \nonumber
 & = & {3 N_e \over x^3} \left[ \sin x - x \cos x\right ] = 3N_e\ \left( {\pi \over 2x^3}\right )^{1/2} \ J_{3/2}(x)
\end{eqnarray}
where $x = \kappa a$, and $J_{3/2}(x)$ is the Bessel function of order $n=3/2$.  The scattering amplitude is proportional to $f^2(\theta) \propto N_e^2$, where $N_e$ the total number of free electrons in the grain. The scattering amplitude for the sphere is the 3-dimensional analog of the scattering amplitude for a circular aperture (Airy disk) which has a diffraction pattern proportional to $J_1^2(x)/x^2.$ 

 \begin{figure}[h]
\begin{center}
\end{center}
\caption{The scattering geometry of an X-rays from a spherical dust grain. Details in text.}
\end{figure}
  
 When the assumptions of the Rayleigh-Gans approximation are not fulfilled, for example for X-rays emitted in the {\it ROSAT} $\onequarter$ and $\threequarters$ keV energy band, the scattering must be calculated using detailed Mie calculations, which are valid even for particles with $a \gg \lambda$ (Smith \& Dwek 1998). The scattering cross section exhibits then a less than quadratic dependence on $N_e$.
 
Regardless of the scattering regime and the numerical procedure used in the calculations, the scattering amplitude is always strongly peaked in the forward direction. This effect is shown in Figure 4, which depicts the polar scattering diagram for 2.9 keV X-rays incident on 0.001, 0.01, 0.1, and 1.0 $\micron$ silicate grains. Figure 5 shows the dependency of the scattering phase function,  defined as $\sigma(\theta)/\pi a^2$, versus scattering angle for 0.01 and 0.1 $\micron$ graphite grains. The figure shows that the scattering phase function is brightest at $\theta$ = 0, steeply declining at angles $ > \lambda/a$. 
 
\subsection{The Collective Effects of Dust Particles on the Halo Profile and Intensity}
Having summarized the size dependent scattering properties of individual dust particles, we will now examine the collective effect of a population of dust particles along the LOS to the source.  The intensity and spatial profile of X-ray halo will depend microscopically on the size distribution and properties of the individual dust grains, and macroscopically on the morphology of the ISM. 

\subsubsection{\bf Clumpy ISM.}
Consider first the formation of a halo in a clumpy ISM. Figure 6a compares the scattering geometry for two single dust clouds located at distances $d_1$ and $d_2$ in front of an X-ray source $S$ as viewed by an observer located at distance $D$ from the source. For sake of clarity, the scattering angles $\theta_1$ and $\theta_2$ are highly exaggerated. The X-ray flux incident on Cloud~1 is larger than that incident on Cloud~2 ($d_1 < d_2$), so Cloud~1 will have a higher surface brightness than Cloud~2. However, the scattering angle $\theta_1$ is larger than $\theta_2$, which can further affect the relative surface brightness of the clouds. To illustrate this effect we will assume that the population of dust particles in each of the two clouds has a bimodal size distribution sharply peaked at radii $a$ = 0.01 and 0.1 $\micron$. If Cloud~1 is sufficiently close to the LOS so that $\theta_1 < \ 200 \arcsec$, then both size populations in each cloud will contribute to the scattering (see Figure 5). However, if $\theta_1 > 200 \arcsec$, the contribution of the population of 0.1 $\micron$ dust particles to the scattering from Cloud 1 will drop dramatically. However, the same population of dust particles may still contribute to the scattering intensity from Cloud~2, provided that $\theta_2 < 200  \arcsec$. This simplified scenario illustrates the fact that: {\it in a clumpy ISM, one can control the intensity and spatial profile of a halo by adjusting the grain size distribution {\it and} the distance of the dust from the source}. 

\begin{figure}[h]
\begin{center}
\end{center}
\caption{The scattering phase function for 2.9 keV X-rays incident on 0.01 and 0.1 $\micron$ graphite particles.}
\end{figure}

\begin{figure}[h]
\begin{center}
\end{center}
\caption{The scattering phase function for 2.9 keV X-rays incident on 0.01 and 0.1 $\micron$ graphite particles for which $\lambda/a$ = $7,400 \arcsec$ and $740\arcsec$, respectively.}
\end{figure}

\subsubsection{\bf Uniform ISM.}
 In an ISM in which the dust is uniformly distributed along the LOS, all scattering angles are possible (see Figure 6b). Without the flexibility of biasing the scattering to a set of discrete angles, one can now only affect the intensity of the halo at, for example, small scattering angles by changing the abundance of small angle scatterers, that is, by adjusting the population of large dust particles. 

\begin{figure}[h]
\caption{Halo geometry for a clumpy ISM (left panel), and a uniform ISM (right panel).}
\end{figure}

\subsubsection{\bf Summary of Factors Determining the Halo Intensity and Profile.}
To summarize, the main factors that play a role in the formation of a halo are: \\ (1) {\it Source properties:} The energy spectrum of the X-ray source affects the halo properties in broad--band X-ray observations. Source variability will cause correlated spatial variations in the intensity and profile, allowing for distance determination and dust distribution to the X-ray source (Alcock \& Hatchett 1978, Predehl et al. 2000). \\  (2) {\it ISM properties:} The distribution of the dust along the LOS, and the total dust column density constrain the scattering geometry. \\ (3) {\it Dust properties:} The dust composition (heavy elements can bind electrons more strongly that light ones - a weak effect), the density of the dust grains, and the grain size distribution are the most fundamental factors determining the halo profile.

\section{Constraining Dust Models With X-Ray Halo Observations}

 \subsection{Nova Cygni 1992}
Nova Cygni 1992 was observed with the {\it ROSAT} satellite, which revealed the existence of an extensive halo surrounding a bright source of soft X-rays in the 0.4 keV energy band. 
The interstellar reddening to this source has been estimated to be between $E(B-V)$ = 0.19 and 0.31, corresponding to an H-column density of $N_{\rm H}$ = 1.1 and 1.8 $\times 10^{21}$ cm$^{-2}$,  respectively (Mathis et al. 1995). The source coordinates are \{l, b\} = \{89.13$\deg$, 7.82$\deg$\}, and its most accepted distance is about 3.2 kpc. The reddening provides an integral constraint on the grain size distribution weighted towards small grains, whereas the X-ray scattering can constrain the large end of the grain size distribution, where $E(B-V)$ is largely flat. 
Simultaneously fitting the observed reddening, halo intensity, and abundance constraints is a delicate balancing act of having enough dust to produce the reddening without overproducing the halo.

Mathis et al. (1995) found that reproducing the observed optical extinction without overproducing the halo intensity, requires the dust to be composed of silicate and amorphous carbon, with voids making up a significant fraction of the volume of the large ($> 0.1 \micron$) grains. Fluffy grains scatter X-rays less efficiently than solid grains of the same size, and can still produce the observed optical extinction.

Smith \& Dwek (1998) challenged the result that fluffy grains are required to simultaneously fit the extinction and halo towards Nova Cygni 1992. Mathis et al. (1995) used the Rayleigh-Gans approximation to calculate the X-ray scattering. This approximation breaks down at the energy of the {\it ROSAT} observations. Detailed Mie calculations show that the X-ray scattering is less efficient than that predicted by the Rayleigh-Gans approximation, and Smith \& Dwek succeeded in reproducing the reddening and the general halo profile with an MRN mixture of bare silicate and graphite grains.

Observational evidence for the presence of large ($> 1 \micron$) interstellar dust grains entering the solar system, summarized by Frisch et al. (1999), prompted Witt, Smith, \& Dwek (2001) to reanalyze the Nova Cygni 1992 halo in search for evidence for the prevalence of such large dust particles in the general ISM. Adopting a uniform distribution of dust along the LOS, an H-column density of $2.1 \times 10^{21}$ cm$^{-2}$, and a reddening value of $E(B-V)$ = 0.36, they found that an MRN size distribution, extended to radii larger than 2 $\micron$ could fit the X-ray halo without violating solar abundance constraints. Their fit was more detailed than the one by Smith \& Dwek (1998), and driven by the halo intensity at the smallest scattering angle in their data of $\sim 100 \arcsec$. Most of the dust mass in the extended MRN distribution, XMRN, is concentrated in the large grains, enhancing the halo intensity at small scattering angles while weakening it at the larger ones. The problem with the XMRN distribution is that it produced a large visual optical depth, leading to an extinction curve with $R_V \equiv A_V/E(B-V) = 6.1$, twice the  value for the average ISM extinction of 3.1. This would suggest a peculiar size distribution along the Nova Cygni 1992 LOS, different from the average ISM. Equivalently, the large extinction produced an $A_V/N_{\rm H}$ value of $1 \times 10^{-22}$ cm$^2$, twice the canonical ISM value. 

Draine \& Tan (2003) reanalyzed the X-ray halo around Nova Cygni 1992 for six different epochs. They presented halo data down to a scattering angle of $\sim 55 \arcsec$, providing stronger constraints on the population of large dust particles, or the LOS distribution of the dust particles. Standard LD01 dust uniformly distributed along the LOS underproduced the halo at small scattering angles, while overproducing it at larger ones (see Figure 15 in their paper). The halo intensity could be increased at small scattering angles by the addition of larger grains in the WD01 size distribution without affecting the halo intensity at large scattering angles. However, they preferred the dust model consisting of LD03 dust with an WD01 size distribution in which 30\% of the dust is in one cloud located close to the source at 0.05 of the source distance to the observer. The model could reproduce the average Galactic extinction with $R_V$ = 3.1, provided that $E(B-V) \approx 0.20$, leading to a H-column density of $N_{\rm H} \approx 1.1 \times 10^{21}$ cm$^{-2}$ for a canonical $A_V/N_{\rm H}$ value of $\approx 5.5 \times 10^{-22}$ cm$^2$. However, such a cloud would have to be at a height of $\sim$ 410~pc above the plane for a source distance of 3.2~kpc. This  height is significantly larger than the $\sim$ 130~pc scale height of the Galactic dust layer (Cox 1999). Draine \& Tan (2003) place the source at a shorter distance of 2.1~kpc, which will still put the cloud at a height of $\sim$ 270~pc above the plane. 

\subsection{The X-Ray Binary GX 13+1}
The high angular resolution of the {\it Chandra} X-ray telescope makes it an ideal tool for studying X-ray halos. Smith, Edgar, \& Shafer (2002, SES02) presented halo observations of GX~13+1 at energies of 2.1, 2.9, and 3.7 keV. In addition to the halo properties, an important constraint on dust models is  the total H-column density to the source, estimated to be $N_{\rm H} = 2.9\pm0.1\ \times 10^{22}$ cm$^{-2}$ (see SES02 for references). SES02 found that the halo could be fitted by an MRN or a WD01 grain size distribution. The observations ruled out fluffy dust models, as well as the XMRN model used by Witt, Smith, \& Dwek (2001). 

The {\it Chandra} observations of GX~13+1 provide a useful means for discriminating between the different dust models of ZDA03.  Figure 7 shows the fit to the 2.9 keV halo intensity profile for a dust model consisting of PAHs and bare silicate and graphite grains (BARE-GR-S), uniformly distributed along the LOS. Similarly good fits were obtained with the same dust model for the halo intensity profile at 2.1 and 3.7 keV. 

\begin{figure}
\begin{center}
\end{center}
\caption{The GX~13+1 X-ray halo profile produced by the BARE-AC-S interstellar dust model (top panel), and by the COMP-AC-S interstellar dust model (lower panel).}
\end{figure}

%
\begin{figure}
\begin{center}
\end{center}
\caption{The scattering phase function of the various dust constituents for the BARE-AC-S dust model (left panel), and for the COMP-AC-S dust model (right panel).}
\end{figure}

Figure 7 also shows that the composite dust model COMP-AC-S provides an equally good fit as the BARE-GR-S model to the halo profile. However, it requires a H-column density of $4.1 \times 10^{22}$ cm$^{-2}$, significantly above the observational value.
The reason for this large column density can be understood by examining the scattering phase function of the different dust constituents in each of the two dust model (Figure 8). COMP dust models contain a population of composite particles which generally comprise larger grains than bare particles. Their scattering phase function exhibits a prominent rise at small scattering angles so that the halo intensity is dominated by composite particles for $\theta < 100 \arcsec$. However, a significant fraction of their volume contains voids, so that their scattering efficiency is significantly reduced compared to same--sized bare particles. In order to fit the halo intensity with such a population of inefficient scatterers their total abundance needs to be increased. However, the ISM abundance constraints prohibit such change, so the only way of fitting the halo with composite particles without violating the abundance constraints is to  increase the H-column density as well. But this quantity is also constrained, by X-ray observations. The overall result is that the halo around GX~13+1 cannot be reproduced by COMP models. This example illustrates the need for reliable H-column density measurement in order to discriminate between otherwise viable dust models. Color excess ratios toward the X-ray sources will also be extremely useful additional constraints. 

\section{Summary}
Dust models depend on the observational constraints that are explicitly included in the fitting procedure, and whether the procedure constrains the grain size distribution to be of a given functional form.
Recently, Zubko, Dwek, \& Arendt (2003) have constructed dust models explicitly constrained by the average interstellar extinction, diffuse IR emission, and ISM abundance constraints, without an a priori functional form for the grain size distribution. Their results show that these constraints are insufficient
for identifying a unique dust model. The radial intensity distribution and spectrum
of X-ray scattering halos provide additional constraints that help characterize the composition and relative numbers of the larger dust
grains. However, since the extinction, IR emission, and abundance constraints represent averages of many different lines of sight, a large observational data base of X-ray halos is needed to render X-ray scattering an equally relevant constraint.

{\bf Acknowledgments} \\
We thank Adolf Witt and the Scientific Organizing Committee for creating a very stimulating conference in a very beautiful and invigorating setting. This work was supported by NASA's Astrophysics Theory Program NRA 99-OSS-01.

\end{document}